# Single-domain perpendicular magnetization induced by the coherent O 2$p$-Ru 4$d$ hybridized state in an ultra-high-quality SrRuO$_3$ film

Yuki K. Wakabayashi,[1,*] Masaki Kobayashi,[2,3,†] Yukiharu Takeda,[4] Kosuke Takiguchi,[1] Hiroshi Irie,[1] Shin-ichi Fujimori,[4] Takahito Takeda,[3] Ryo Okano,[3] Yoshiharu Krockenberger,[1] Yoshitaka Taniyasu,[1] and Hideki Yamamoto[1]

[1]*NTT Basic Research Laboratories, NTT Corporation, Atsugi, Kanagawa 243-0198, Japan*
[2]*Center for Spintronics Research Network, The University of Tokyo, 7-3-1 Hongo, Bunkyo-ku, Tokyo 113-8656, Japan*
[3]*Department of Electrical Engineering and Information Systems, The University of Tokyo, Bunkyo, Tokyo 113-8656, Japan*
[4]*Materials Sciences Research Center, Japan Atomic Energy Agency, Sayo-gun, Hyogo 679-5148, Japan*

[*]Corresponding author: yuuki.wakabayashi.we@hco.ntt.co.jp
[†]Corresponding author: masaki.kobayashi@ee.t.u-tokyo.ac.jp

Abstract

We investigated the Ru 4$d$ and O 2$p$ electronic structure and magnetic properties of an ultra-high-quality SrRuO$_3$ film on SrTiO$_3$ grown by machine-learning-assisted molecular beam epitaxy. The high itinerancy and long quantum lifetimes of the quasiparticles in the Ru 4$d$ $t_{2g}$-O 2$p$ hybridized valence band are confirmed by observing the prominent well-screened peak in the Ru 3$d$ core-level photoemission spectrum, the coherent peak near the Fermi energy in the valence band spectrum, and quantum oscillations in the resistivity. The element-specific magnetic properties and the hybridization between the Ru 4$d$ and O 2$p$ orbitals were characterized by Ru $M_{2,3}$-edge and O $K$-edge soft X-ray absorption spectroscopy and X-ray magnetic circular dichroism measurements. The ultra-high-quality SrRuO$_3$ film with the residual resistivity ratio of 86 shows the large orbital magnetic moment of oxygen ions induced by the strong orbital hybridization of the O 2$p$ states with the spin-polarized Ru 4$d$ $t_{2g}$ states. The film also shows single-domain perpendicular magnetization with an almost ideal remanent magnetization ratio of 0.97. These results provide detailed insights into the relevance between orbital hybridization and the perpendicular magnetic anisotropy in SrRuO$_3$/SrTiO$_3$ systems.

## I. INTRODUCTION

The itinerant 4$d$ ferromagnetic perovskite SrRuO$_3$ [bulk Curie temperature ($T_C$) = 160 K] has been studied extensively for many decades because of the unique nature of its ferromagnetism, metallicity, chemical stability, and compatibility with other perovskite-structured oxides [1–21]. Unlike many perovskite transition-metal oxides, the spin-polarized Ru 4$d$ orbitals hybridized with the O 2$p$ orbitals in SrRuO$_3$ have the itinerant character despite the strong electron correlation [8]. Therefore, SrRuO$_3$ is widely used as a ferromagnetic metal electrode in oxide electronic and spintronic devices consisting of perovskite layers. In addition, the perpendicular magnetic anisotropy induced by the compressive strain [7,8,20] is beneficial for scalability and the reduction of power consumption in spintronic devices, such as magnetic random access memory, and thus SrRuO$_3$-based all-oxide spintronic devices have been investigated [15,16,22].

SrRuO$_3$/SrTiO$_3$ is the first oxide heterostructure in which perpendicular magnetic anisotropy was discovered [23], and it has been a model system for understanding it in metallic oxides [24,25]. It has been generally considered that perpendicular magnetic anisotropy arises from magnetocrystalline anisotropy caused by spin-orbit interactions. Bruno has demonstrated that the magnetocrystalline anisotropy energy is proportional to the difference in the orbital magnetic moment between the perpendicular and in-plane directions [26]. Indeed, a large orbital magnetic moment perpendicular to the films (0.08-0.1 $\mu_B$/Ru) below $T_C$ has been reported in SrRuO$_3$ films on SrTiO$_3$ [24,25]. Recently, a polarized neutron diffraction experiment revealed the unexpected large magnetic moment of oxygen, which contributes 30% of the total magnetization in the case of bulk SrRuO$_3$ [27]. Besides, Jeong $et$ $al.$ have reported a metal-insulator transition caused by a weakened Ru 4$d$ $t_{2g}$-O 2$p$ hybridization near SrRuO$_3$/SrTiO$_3$ interfaces [28]. These results highlight the importance of the O 2$p$ states for understanding the perpendicular magnetic anisotropy and electronic structures in SrRuO$_3$ films, and the issues remain controversial. Since the controversy may partly arise from variation in sample quality in previous experiments, it is vitally important to investigate the element-specific electronic structures and magnetic properties using very high-quality SrRuO$_3$/SrTiO$_3$ films.

The residual resistivity ratio (RRR), defined as the ratio of the longitudinal resistivity $\rho$ at 300 K [$\rho$(300 K)] and $T$→0 K [$\rho$($T$→0 K)] ($T$: temperature), is an excellent measure to gauge the purity of a metallic system: the quality of single-crystalline SrRuO$_3$ thin films. High RRR values are essential for exploring intrinsic electronic states. In particular, SrRuO$_3$ thin films with RRR values above 20 have enabled observations of dispersive quasiparticle peaks near the Fermi level ($E_F$) by angle-resolved photoemission spectroscopy [13] as well as quantum oscillations of Weyl fermions [18,29,30] and trivial Ru 4$d$ electrons [14] via electrical resistivity [i.e., Shubnikov-de Haas (SdH) oscillations] measurements. Thus, ultra-high-quality SrRuO$_3$/SrTiO$_3$ films with very high RRR values could provide promising opportunities to comprehend the perpendicular magnetic anisotropy in this system.

In this study, we investigated the Ru 4$d$ and O 2$p$ electronic structure and magnetic properties using the ultra-high-quality SrRuO$_3$ films grown on SrTiO$_3$ substrates by soft X-ray photoemission spectroscopy (SX-PES), soft X-ray absorption spectroscopy (XAS), and X-ray magnetic circular dichroism (XMCD). To characterize the element-specific magnetic properties and the hybridization strength between the Ru 4*d* and O 2*p* orbitals, both Ru $M_{2,3}$-edge and O *K*-edge absorptions were used in XAS and XMCD measurements. To form SrRuO$_3$ with quality exceeding current levels, we employed our recently developed machine-learning-assisted molecular beam epitaxy (MBE) [31]. The ultra-high-quality SrRuO$_3$ film, having the highest RRR of 86, allows to access intrinsic properties of SrRuO$_3$. We found large orbital magnetic moments of oxygen ions accompanied by strong orbital hybridization of the O 2*p* states with the spin-polarized Ru 4*d* $t_{2g}$ states. The film also shows single-domain perpendicular magnetization with an almost ideal remanent magnetization ratio of 0.97.

## II. EXPERIMENT

We grew high-quality epitaxial SrRuO$_3$ films with a thickness of 63 nm on the (001) SrTiO$_3$ substrates in a custom-designed MBE setup equipped with multiple e-beam evaporators for Sr and Ru. The growth parameters were optimized by Bayesian optimization, a machine learning technique for parameter optimization [31–33], with which we achieved RRR = 86. The growth temperature was 772°C. We precisely controlled the elemental fluxes, even for elements with high melting points, e.g., Ru (2250°C), by monitoring the flux rates with an electron-impact-emission-spectroscopy sensor, which were fed back to the power supplies for the e-beam evaporators. The Ru and Sr fluxes were 0.365 and 0.980 Å/s, respectively, corresponding to Ru-rich conditions. Excessive Ru is known to be desorbed from the growth surface by forming volatile species such as RuO$_4$ and RuO$_3$ under an oxidizing atmosphere, leading to stoichiometric films [10,11]. The growth rate of 1.05 Å/s was deduced from the thickness calibration of the film using cross-sectional scanning transmission electron microscopy (STEM). This growth rate agrees very well with the value of 1.08 Å/s estimated from the flux rate of Sr, confirming the accuracy of the film thickness and thus the absolute values of the resistivity, conductivity, and magnetic moment. The oxidation during growth was carried out with a mixture of ozone (O$_3$) and O$_2$ gas (~15% O$_3$ + 85% O$_2$), which was introduced at a flow rate of ~2 sccm through an alumina nozzle pointed at the substrate. The nozzle-to-substrate distance was 15 mm. Further information about the MBE setup and preparation of the substrates is available elsewhere [34–36].

For the magnetotransport measurements, we first deposited Ag electrodes on a SrRuO$_3$ surface. Then, we patterned the samples into 200 × 350 μm$^2$ Hall bar structures by photolithography and Ar ion milling. Resistivity was measured using the four-probe method at 100 μA in a Physical Property Measurement System (PPMS) DynaCool sample chamber equipped with a rotating sample stage. Low-noise measurements were performed by an AC analog lock-in technique below 1 K, for which the sample was cooled down in a $^3$He–$^4$He dilution refrigerator.

The sample was transferred in air to the helical undulator beamline BL23SU of SPring-8 [37-40] to perform the SX-PES, XAS, and XMCD measurements. The monochromator resolution $E/\Delta E$ was about 10,000. For the XMCD measurements, absorption spectra for circularly polarized X rays with the photon helicity parallel ($\mu^+$) and antiparallel ($\mu^-$) to the spin polarization were obtained by reversing the photon helicity at each photon energy $h\nu$ and recorded in the total-electron-yield (TEY) mode.

The $\mu^+$ and $\mu^-$ spectra at the Ru $M_{2,3}$ edges and O $K$ edge were taken for both positive and negative applied magnetic fields and averaged to eliminate spurious dichroic signals. External magnetic fields were applied perpendicular to the sample surface ([001] direction of the SrTiO$_3$ substrate). The sample temperature was varied between 6.5 and 200 K. For estimation of the integrated values of the XAS spectra at the Ru $M_{2,3}$ edge, hyperbolic tangent functions were subtracted from the spectra as background. The sample was kept at 20 K under an ultra-high vacuum better than $10^{-8}$ Pa during the PES measurements. The total energy resolution was ~150 meV. The position of the Fermi level ($E_F$) was determined by measuring the Fermi cutoff of evaporated gold in electrical contact with the samples.

**III. RESULTS AND DISCUSSION**

Figures 1(a) and 1(b) show RHEED patterns of the SrRuO$_3$ surface, in which very sharp streaks with Kikuchi lines and higher-order Laue patterns can be clearly seen. This indicates that a high-quality SrRuO$_3$ layer was epitaxially grown in a two-dimensional growth mode. Figures 1(c) and 1(d) show high-angle annular dark-field STEM (HAADF-STEM) and annular bright-field STEM (ABF-STEM) images of the SrRuO$_3$ film. SrRuO$_3$ grew epitaxially on a (001) SrTiO$_3$ substrate with an abrupt substrate/film interface, as expected from the RHEED patterns. The SrRuO$_3$ film is compressively strained because the lattice constant of the SrTiO$_3$ substrate (3.905 Å) is ~0.6% smaller than the pseudocubic bulk lattice constant of SrRuO$_3$ [31].

The resistivity $\rho$ vs. temperature $T$ curve of the SrRuO$_3$ film shows a clear kink at 152 K [Fig. 2(a)]. The kink corresponds to the $T_C$ where the ferromagnetic transition occurs, and spin-dependent scattering is suppressed [8]. With a residual resistivity $\rho(T\rightarrow 0$ K) of 2.18 $\mu\Omega\cdot$cm and an RRR of 86, the SrRuO$_3$ film grown by machine-learning-assisted MBE is superior to those prepared by any other method [7,8]. As shown in the inset in Fig. 1(a), below 20 K, the SrRuO$_3$ film showed a $T^2$ scattering rate ($\rho \propto T^2$) that is expected for a Fermi liquid, in which electron-electron scattering dominates the transport and carriers are described as Landau quasiparticle [8,14,41]. In this Fermi liquid temperature range, quantum lifetimes long enough to observe quantum oscillations are achieved, as evidenced by the observation of the SdH oscillations of Weyl fermions with low frequencies of 25-32 T [Fig. 2(b)] [29] and those of trivial Ru 4$d$-O 2$p$ band electrons with high frequencies of 360 and 3400-3800 T [inset in Fig. 2(b)] [13,14,29] (see also the Supplemental Material [42]).

To elucidate the electronic structure, we performed SX-PES measurements on the SrRuO$_3$ film. Figure 3(a) shows the Ru 3$d$ core-level spectrum of the SrRuO$_3$ film taken with $h\nu$ = 1200 eV. Here, the peaks at 285.6 and 278.6 eV were assigned to C 1$s$ (from air contamination at the surface) and Sr 3$p_{1/2}$ [27,43,44] signals. In the Ru 3$d$ spectrum, the Ru 3$d_{5/2}$ component is composed of a well-screened (WS) peak (281.4 eV) and a poorly-screened (PS) peak (282.9 eV). The WS peak results from the screening of the Ru 3$d$ core-hole state by conduction electrons [43-45]. Notably, our SrRuO$_3$ film prepared by machine-learning-assisted MBE shows a prominent and sharp WS peak, which is in contrast to the case of film grown by pulsed laser deposition, where the WS peak was observed only as a small shoulder of the PS peak [27, 43]. To further scrutinize this point, the spectrum was fitted to a combination of Voigt and asymmetric Gaussian functions. As shown in Fig. 3(b), the fitting well reproduces the Ru $3d_{5/2}$ and Sr $3p$ structures. In accordance with the spectral line shape of the Ru $3d$ spectrum, the WS peak is prominent rather than the PS one. This means that a Ru $3d$ core-hole state created by the photoemission process is well screened by the conduction electrons, which is consistent with the good metallic conductivity of our SrRuO$_3$ film. Moreover, there is a possibility that the Weyl fermions also contribute to this screening. The concentration of conduction electrons is related to the density of states near $E_F$. Figure 3(c) shows the valence band spectrum taken with $h\nu = 600$ eV. The intense coherent peak originating from the Ru $4d$ states near the $E_F$ is observed. These observations of the sharp WS peak in the Ru $3d_{5/2}$ core-level spectrum and the coherent peak at $E_F$ in the valence-band spectrum provide spectral evidence for the good metallicity of the SrRuO$_3$ film.

To get detailed insights into the orbital hybridization and the perpendicular magnetic anisotropy, we carried out XAS and XMCD measurements, which are tools sensitive to the local electronic structure and element-specific magnetic properties in magnetic materials [46-50]. Figure 4(a) shows the Ru $M_{2,3}$ XAS and XMCD spectra of the SrRuO$_3$ film at 6.5 K with a magnetic field of $\mu_0 H = 0$ T. The spectra were measured after the application of $\mu_0 H = 2$ T that is enough high to saturate the magnetization, and hence, the XMCD signals originate from the remanent-spontaneous magnetization. Here, $\mu^+$ and $\mu^-$ denote the absorption coefficients for the photon helicities parallel and antiparallel to the Ru $4d$ majority spin direction, respectively. The absorption peaks at 463 and 485 eV are due to transitions from the Ru $3p_{3/2}$ and $3p_{1/2}$ core levels into the Ru $4d$ band [51]. Other structures located around 476 and 498 eV are attributed to transitions into Ru $5s$ states [24]. The peak position of the $M_3$-edge XMCD is lower than that of the $M_3$-edge XAS, indicating that the XMCD and XAS peaks come from the transitions to the Ru 4d $t_{2g}$ and $e_g$ bands, respectively, and only the Ru $4d$ $t_{2g}$ states near the $E_F$ have spin polarization. These assignments are consistent with our previous density functional theory (DFT) calculations [29], in which the half-metallic Ru $4d$ $t_{2g}$ states cross the $E_F$.

We determined the orbital magnetic moment $m_{\text{orb}}$ and the spin magnetic moment $m_{\text{spin}}$ of the Ru$^{4+}$ $4d$ states using the XMCD sum rules as follows [46-48]:

$$m_{\text{orb}} = -\frac{4(10-n_{4d})}{3r}\int_{M_2+M_3}(\mu^+ - \mu^-)dE,$$

$$m_{\text{spin}} + m_{\text{T}} = -\frac{2(10-n_{4d})}{r}[\int_{M_3}(\mu^+ - \mu^-)dE + 2\int_{M_2}(\mu^+ - \mu^-)dE].$$

Here, $r = \int_{M_2+M_3}(\mu^+ + \mu^-)dE$, and $n_{4d}$ is the number of electrons in $4d$ orbitals, which is assumed to be four. For ions in octahedral symmetry, the magnetic dipole moment $m_{\text{T}}$ is a small number and can be neglected compared to $m_{\text{spin}}$ [52]. Using the XMCD spectra taken with a magnetic field of $\mu_0 H = 2$ T at 6.5 K [Fig. 4(b)], we estimated the $m_{\text{spin}} = 0.85$ $\mu_B$/Ru and $m_{\text{orb}} = 0.07$ $\mu_B$/Ru. The orbital magnetic moment relative to the spin magnetic moment, $m_{\text{orb}}/m_{\text{spin}}$, is 0.08, consistent with the reported value for SrRuO$_3$ films grown on SrTiO$_3$ (001) substrates [24,25]. The total magnetic moment, $M = m_{\text{spin}} + m_{\text{orb}} = 0.92$, is smaller than the saturation magnetization measured with a superconducting quantum interference device magnetometer (1.25 $\mu_B$/Ru) [31]. This slight discrepancy may come from the magnetization of the O $2p$ electrons, which should be induced by orbital hybridization of the O 2p states with the spin-polarized Ru 4d states [53]. As described later, indeed, the substantial orbital magnetic moment of the O 2p states was verified from the O 1s XMCD spectrum.

Figure 5(a) shows the temperature-dependent (6.5 – 200 K) Ru $M_{2,3}$-edge XMCD spectra with a magnetic field $\mu_0 H$ = 0 or 2 T. The XMCD intensities decrease with increasing $T$ and drop above $T_C$ (152 K), reflecting the ferromagnetic to paramagnetic transition. Figure 5(b) shows the XMCD spectra normalized at the Ru $M_2$ edge. Below $T_C$, the normalized XMCD spectra are identical to each other, indicating that $m_{\text{orb}}/m_{\text{spin}}$ stays constant below $T_C$. In contrast, the normalized Ru $M_3$-edge intensity in the paramagnetic state is smaller than those in the ferromagnetic state [Fig. 5(b)]. Using the XMCD spectra at 200 K shown in Fig. 5(c), we estimated the $m_{\text{spin}}$ = 0.037∓0.01 $\mu_B$/Ru and $m_{\text{orb}}$ = 0∓0.01 $\mu_B$/Ru in the paramagnetic state. The total magnetic moment drops above $T_C$, and $m_{\text{orb}}$ is quenched within the accuracy of the measurements, indicating that the Ru$^{4+}$ states are in the low-spin $S$ = 1 [$t_{2g}^4$ (3↑, 1↓)] paramagnetic state ($2\sqrt{S(S+1)}$ = 2.83 $\mu_B$/Ru). This is consistent with the reported experimental effective moment of bulk SrRuO$_3$ above $T_C$ (~2.6 $\mu_B$/Ru) [2,54]. The lower experimental saturation magnetization below $T_C$ has been attributed to electron delocalization associated with itinerancy [25,55].

To clarify the unoccupied electronic states hybridized with the O 2p orbitals, we measured the O 1s XAS and XMCD spectra [Fig. 6]. The O 1s XAS spectra of transition metal oxides represent the unoccupied transition metal 4d and 5s/5p states, as well as the other conduction-band states via the hybridization with the unoccupied O 2p states [56]. The absorption peak at 529 eV comes from transitions into the Ru 4d $t_{2g}$ states, and the transitions to the Ru 4d $e_g$ states appear in the energy range of 530–534.5 eV [57]. The transitions to the Sr 4d states and the Ru 5s states are observed in the range of 534.5–540 eV and 540–546 eV [43,57,58], respectively. The energy difference between the Ru 4d $t_{2g}$ peak and the Ru 5s peak (~14.5 eV) is consistent with that in the Ru $M_{2,3}$-edge XAS and XMCD spectra (~14 eV) [Fig. 4(a)]. The O 1s XAS spectrum of the ultra-high-quality SrRuO$_3$ film exhibits sharper peaks than other SrRuO$_3$ films on SrTiO$_3$ [28,43] and bulk SrRuO$_3$ [56,57,58] with similar overall features. Notably, the intensity ratio of the Ru 4d $t_{2g}$ peak to the Sr 4d peak (1.55) is larger than those in other SrRuO$_3$ films on SrTiO$_3$ and bulk SrRuO$_3$ (0.8-1.3) [28,43,56-58]. The smaller Ru 4d $t_{2g}$ peak in previous studies may come from the less itinerant nature (shorter lifetimes) of the quasiparticles in the hybridized O 2p-Ru 4d $t_{2g}$ states crossing the $E_F$, as a consequence of, for instance, disorder-induced localization. The localization comes from hybridization between different basis states caused by the disorder when seen as a perturbation to the Hamiltonian in the zero-disorder limit [59].

The substantial orbital magnetic moment of the O 2p states induced by the hybridization with the Ru 4d $t_{2g}$ states was observed from the O 1s XMCD spectrum (Fig. 6). The sizeable negative XMCD structure was observed only in the energy range corresponding to Ru 4d $t_{2g}$, indicating that the influence of the magnetization due to hybridizations with the states other than Ru 4d $t_{2g}$ is negligibly small. This is consistent with our previous DFT calculations [29], in which the half-metallic bands crossing the $E_F$ are formed by the Ru 4d $t_{2g}$ states hybridized with the O 2p states. The O 1s XMCD peak intensity at 529.1 eV divided by the O 1$s$ XAS peak intensity at 529 eV (0.13) is twice as large as that of bulk polycrystalline SrRuO$_3$ (~0.06) [39]. Since the energy-integrated intensity of the O 1$s$ XMCD is proportional to the orbital magnetic moment of the O 2$p$ states [46], the larger O 1$s$ XMCD intensity indicates the large orbital magnetic moment of the O 2$p$ states in the ultra-high-quality SrRuO$_3$ film. According to P. Bruno [26], the larger orbital magnetic moment perpendicular to the film should lead to perpendicular magnetic anisotropy, as confirmed by XMCD for several systems, including Co thin films sandwiched by Au(111) [60] and FePt [61]. Thus, the large orbital magnetic moments of the hybridized O 2$p$-Ru 4$d$ $t_{2g}$ states in ultra-high-quality SrRuO$_3$ films will lead to the strong perpendicular magnetic anisotropy. Indeed, an epitaxial SrRuO$_3$ film, whose RRR is over 50, on SrTiO$_3$ showed the largest perpendicular magnetic anisotropy among SrRuO$_3$ films on SrTiO$_3$ reported at the time [31].

Figure 7 shows XMCD-$H$ curves measured at the Ru $M_3$ edge (462.4 eV) for our ultra-high-quality SrRuO$_3$ film at 6.5 K. The vertical axis of the XMCD intensity at 2 T has been scaled so that it represents the sum of the total magnetic moment $M = m_{\text{spin}} + m_{\text{orb}}$ of the Ru ions estimated from Fig. 4(b). The rectangular hysteresis with the small coercive field $H_\text{C}$ of ~0.02 T means that the easy direction of magnetization is perpendicular to the film surface, as is usually the case with compressively strained SrRuO$_3$ films on SrTiO$_3$ substrates [8,29,31]. The $H_\text{C}$ of ~0.02 T is smaller than those previously reported for SrRuO$_3$ films ($H_\text{C}$ > 0.1 T) [8,29,31]. Since the magnetic domains tend to be pinned by grain boundaries and other defects, the small $H_\text{C}$ stems from the higher crystallinity of our film. Notably, the ratio of remanent magnetization to saturation magnetization, estimated by the $M$ values at 0 and ±2 T in Fig. 7, is almost an ideal value of 0.97, indicating the single-domain perpendicular magnetization. To our knowledge, this is the highest value reported for all oxides [62-64]. We note that, for accurate determination of the remanent magnetization ratio of magnetic thin films, XMCD is adequate because it is free from the diamagnetic signal from the substrate. The nearly ideal perpendicular magnetization mentioned above, which stems from the large orbital magnetic moment of the O 2$p$-Ru 4$d$ hybridized states as well as less grain boundary and defect densities, is beneficial for spintronics applications; with larger perpendicular magnetization, the magnetic configuration is thermally more stable, and the spin-transfer switching current is lower [65,66].

**IV. CONCLUSIONS**

We have investigated the Ru 4$d$ and O 2$p$ electronic structure and magnetic properties of the ultra-high-quality SrRuO$_3$ film on SrTiO$_3$ grown by machine-learning-assisted MBE. The thusly prepared SrRuO$_3$ film, with a residual resistivity $\rho$(T→0 K) of 2.18 μΩ·cm and a RRR of 86, is superior to those obtained by any other method, allowing access the intrinsic properties of SrRuO$_3$. We observed the prominent well-screened peak in the Ru 3$d_{5/2}$ core-level spectrum and the coherent Ru 4$d$ $t_{2g}$ peak at $E_\text{F}$ by SX-PES. Together with quantum oscillations in the resistivity, the highly itinerant nature (long quantum lifetime) of the quasiparticles in the Ru 4$d$ $t_{2g}$-O 2$p$ hybridized bands crossing the $E_\text{F}$ is confirmed. We also revealed large orbital magnetic moments of oxygen ions and the strong orbital hybridization of the O 2$p$ states with the spin-polarized

Ru $4d$ $t_{2g}$ states. The O $2p$-Ru $4d$ $t_{2g}$ hybridization in the ultra-high-quality SrRuO$_3$ film is more significant than those in other SrRuO$_3$ films on SrTiO$_3$ and in bulk SrRuO$_3$ [28,43,56-58], and this strong O $2p$-Ru $4d$ $t_{2g}$ hybridization is responsible for the high itinerancy of the quasiparticles at around the Fermi level. The ultra-high-quality SrRuO$_3$ film shows single-domain perpendicular magnetization with an almost ideal remanent magnetization ratio of 0.97. These results provide important insights into the relevance between the orbital hybridization and perpendicular magnetic anisotropy in SrRuO$_3$/SrTiO$_3$ systems and for applying SrRuO$_3$ as a metallic ferromagnetic oxide electrode for hetero-epitaxially grown spintronic devices.

**ACKNOWLEDGMENTS**
This work was partially supported by JST-CREST (JPMJCR18T5) and the Spintronics Research Network of Japan (Spin-RNJ). This work was performed under the Shared Use Program of Japan Atomic Energy Agency (JAEA) Facilities (Proposal No. 2020A-E18) supported by JAEA Advanced Characterization Nanotechnology Platform as a program of "Nanotechnology Platform" of the Ministry of Education, Culture, Sports, Science and Technology (MEXT) (Proposal No. JPMXP09A20AE0018). The experiment at SPring-8 was approved by the Japan Synchrotron Radiation Research Institute (JASRI) Proposal Review Committee (Proposal No. 2020A3841).

**AUTHORS' CONTRIBUTIONS**
Y.K.W. conceived the idea, designed the experiments, and directed and supervised the project. M.K. and Y.K.W. planned the synchrotron experiments. Y.K.W. and Y.K. grew the samples. Y.K.W. carried out the sample characterizations. K.T., Y.K.W., and H.I. carried out the magnetotransport measurements. Y.K.W., M.K., Y.Tak., T.T., and R.O. carried out the XMCD measurements. M.K. and S.-I.F. carried out the X-ray PES measurements. Y.K.W. and M.K. analyzed and interpreted the data. Y.K.W. wrote the paper with input from all authors.

**DATA AVAILABILITY**
Data that support the findings of this study are available from the corresponding author upon reasonable request.

**Figures and figure captions**

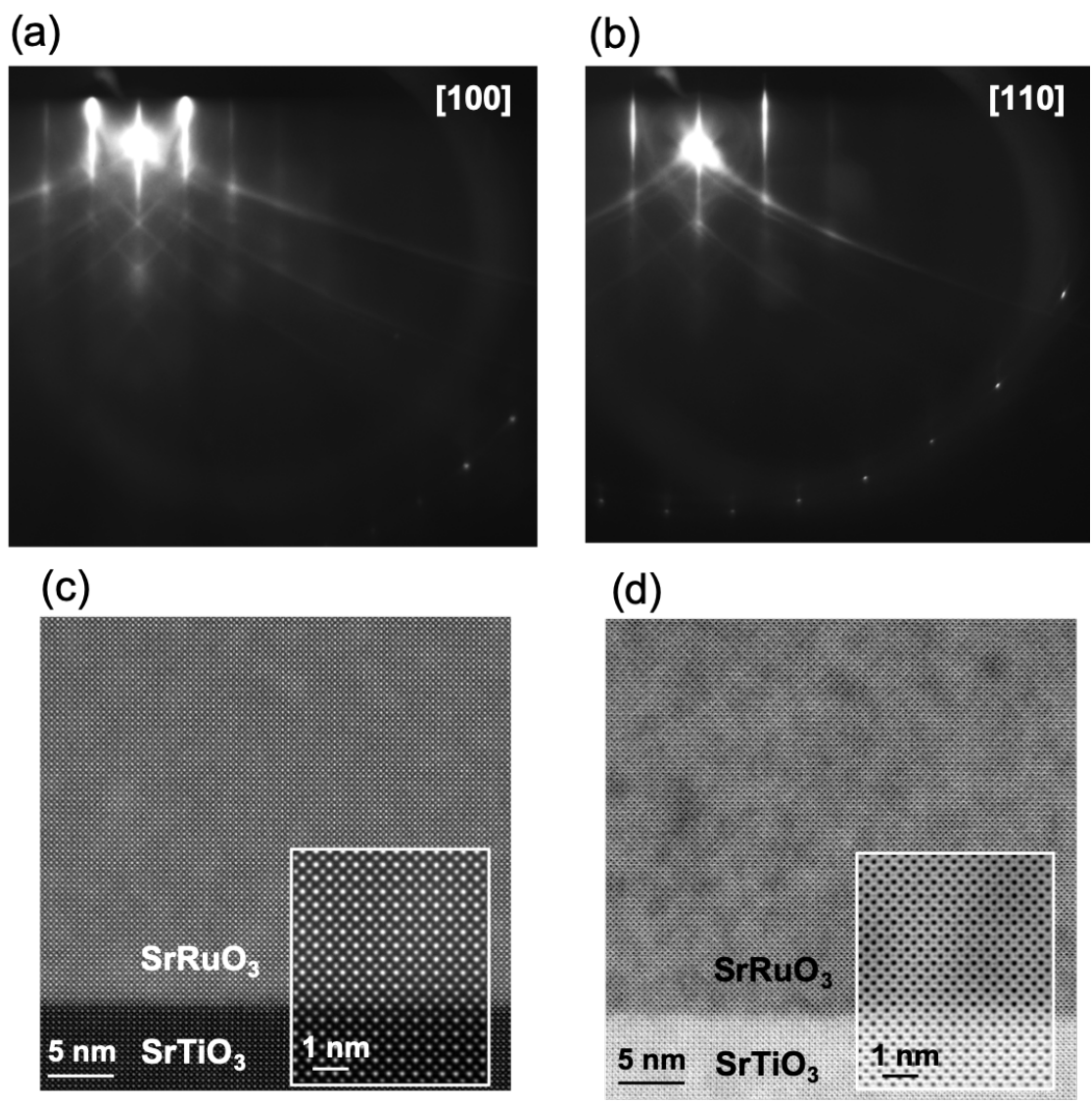

FIG. 1. RHEED patterns of a SrRuO$_3$ film taken along the (a) [100] and (b) [110] axes of the SrTiO$_3$ substrates. (c) HAADF-STEM and (d) ABF-STEM images of a SrRuO$_3$ film taken along the [100] axis of the SrTiO$_3$ substrates. The insets in (c) and (d) are magnified images near the interface.

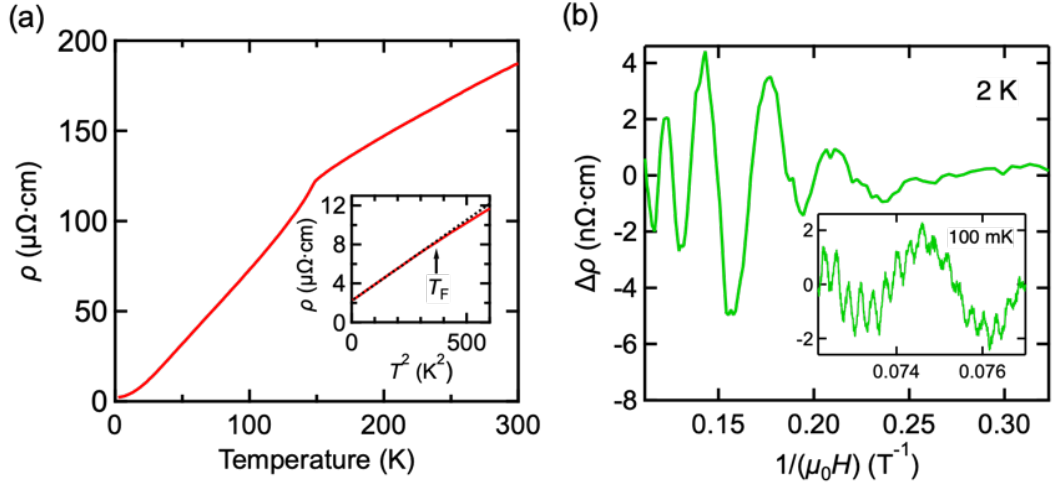

FIG. 2. (a) Temperature dependence of resistivity $\rho$ of the SrRuO$_3$ film. The inset in (a) is a $\rho$ versus $T^2$ plot with a linear fitting (black dashed line). We defined the Fermi liquid temperature range as the temperature range where the experimental $\rho_{xx}$ and the linear fitting line in $\rho_{xx}$ vs. $T^2$ are close enough to each other (< 0.1 μΩ·cm). (b) SdH oscillation measured at 2 K with $\mu_0H$ (3 T < $\mu_0H$ < 9 T) applied in the out-of-plane [001] direction of the SrTiO$_3$ substrate for the SrRuO$_3$ film. The inset shows SdH oscillation observed at 0.1 K with $\mu_0H$ (13 T < $\mu_0H$ < 14 T).

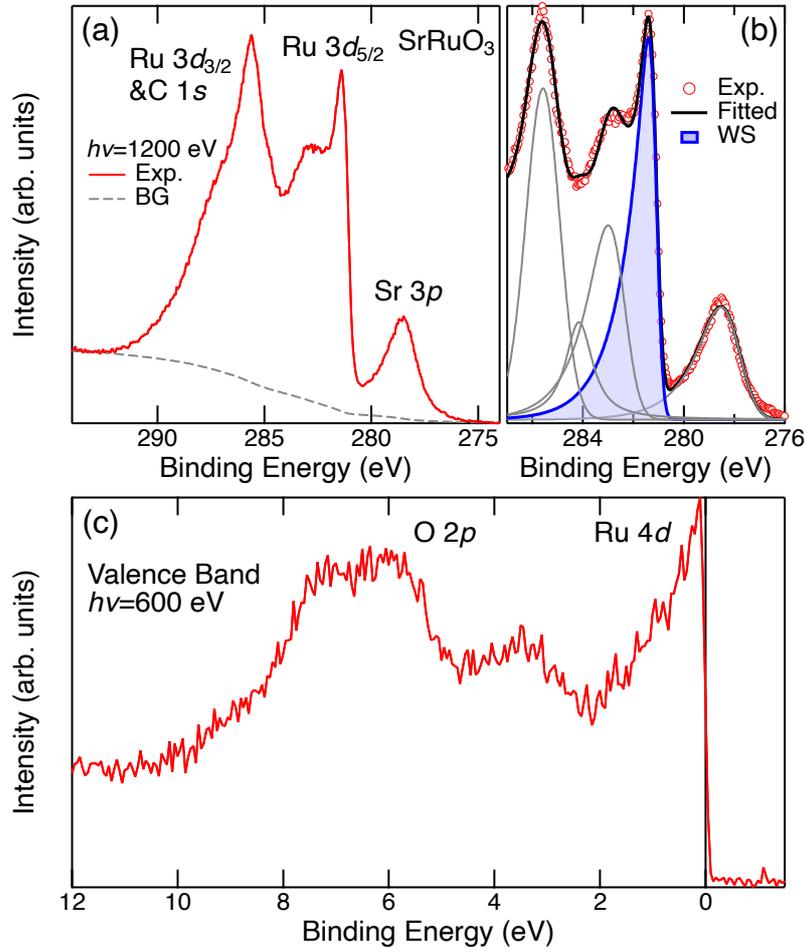

FIG. 3. SX-PES spectra of the SrRuO$_3$ film. (a) Ru 3$d$ and Sr 3$p$ core-level spectrum taken with the photon energy ($h\nu$) of 1250 eV at 20 K. (b) Peak fitting for the Ru 3$d$ spectrum. WS denotes the well-screened peak. Fitting functions are Voigt and asymmetric Gaussian. (c) Valence band spectra taken with $h\nu$ of 600 eV at 20 K.

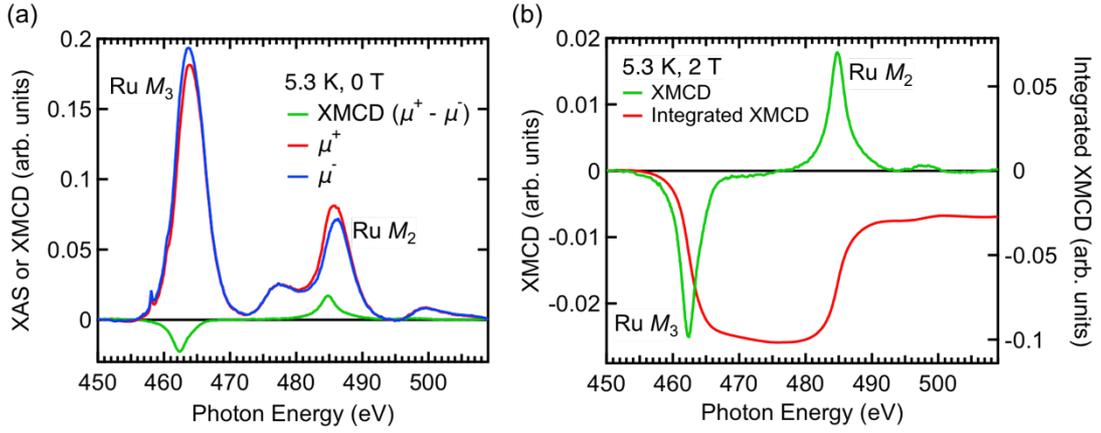

FIG. 4. (a) Ru $M_{2,3}$ edge XAS and XMCD spectra for the SrRuO$_3$ film at 6.5 K with a magnetic field $\mu_0 H = 0$ T applied perpendicular to the film surface. The spectra were measured after the application of $\mu_0 H = 2$ T. (b) XMCD spectra and integrated XMCD signals from 450 eV taken at 6.5 K with a magnetic field $\mu_0 H = 2$ T applied perpendicular to the film surface.

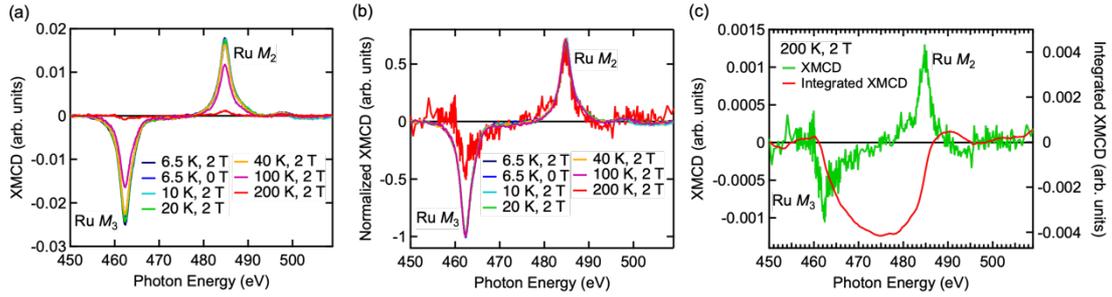

FIG. 5. (a) Ru $M_{2,3}$ edge XMCD spectra for the SrRuO$_3$ film at 6.5 to 200 K with a magnetic field $\mu_0H$ = 0 or 2 T applied perpendicular to the film surface. Here, the zero magnetic field was set after the application of $\mu_0H$ = 2 T. (b) Ru $M_{2,3}$ edge XMCD spectra normalized at 484.7 eV for the SrRuO$_3$ film at 6.5 to 200 K with a magnetic field $\mu_0H$ = 0 or 2 T applied perpendicular to the film surface. Spectra measured at 6.5, 10, 20, 40, and 100 K are almost completely overlapped. (c) XMCD spectra and integrated XMCD signals from 450 eV at 200 K with a magnetic field $\mu_0H$ = 2 T applied perpendicular to the film surface.

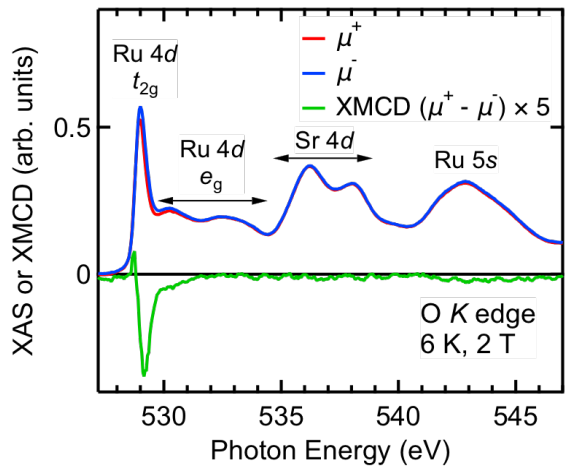

FIG. 6. O *K* edge XAS and XMCD spectra for the SrRuO$_3$ film at 6 K with a magnetic field $\mu_0 H$ = 2 T applied perpendicular to the film surface.

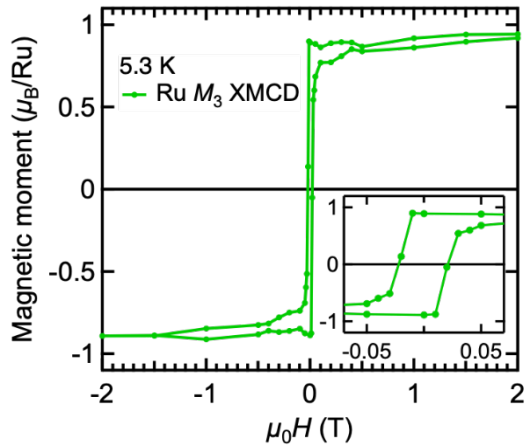

FIG. 7. XMCD-*H* curves measured at the Ru $M_3$ edge (462.4 eV) for the SrRuO$_3$ film at 6.5 K. The vertical axis of the XMCD intensity at 2 T has been scaled so that it represents the sum of the total magnetic moment $M = m_{\text{spin}} + m_{\text{orb}}$ of the Ru ions estimated from Fig. 4(b).

Supplemental Material
**Single-domain perpendicular magnetization induced by the coherent O 2p-Ru 4d hybridized state in an ultra-high-quality SrRuO$_3$ film**

Yuki K. Wakabayashi,[1,*] Masaki Kobayashi,[2,3,†] Yukiharu Takeda,[4] Kosuke Takiguchi,[1] Hiroshi Irie,[1] Shin-ichi Fujimori,[4] Takahito Takeda,[3] Ryo Okano,[3] Yoshiharu Krockenberger,[1] Yoshitaka Taniyasu,[1] and Hideki Yamamoto[1]

[1]*NTT Basic Research Laboratories, NTT Corporation, Atsugi, Kanagawa 243-0198, Japan*
[2]*Center for Spintronics Research Network, The University of Tokyo, 7-3-1 Hongo, Bunkyo-ku, Tokyo 113-8656, Japan*
[3]*Department of Electrical Engineering and Information Systems, The University of Tokyo, Bunkyo, Tokyo 113-8656, Japan*
[4]*Materials Sciences Research Center, Japan Atomic Energy Agency, Sayo-gun, Hyogo 679-5148, Japan*

[*]Corresponding author: yuuki.wakabayashi.we@hco.ntt.co.jp
[†]Corresponding author: masaki.kobayashi@ee.t.u-tokyo.ac.jp

**Shubnikov-de Haas (SdH) oscillations of Weyl fermions and electrons in the trivial Ru 4$d$ bands.**

Figure S1(a) shows the magnetoresistance (MR) ($\rho(\mu_0H)-\rho(0\ \text{T}))/\rho(0\ \text{T})$ at 2 K with $\mu_0H$ (3 T < $\mu_0H$ < 9 T) applied in the out-of-plane [001] direction of the SrTiO$_3$ substrate. The linear positive MR, which is thought to stem from the linear energy dispersion of Weyl nodes, was observed [S1-S3]. Pretreatments of the Shubnikov-de Haas (SdH) oscillation data are crucial for deconvoluting quantum oscillation spectra, since magnetoconductivity data generally contain not only oscillation components but also other magnetoresistive components as background signals [S2,S4]. In particular, SdH oscillations in SrRuO$_3$ are subject to being masked by large non-saturated positive MR [Fig. S1(a)]. Here, we subtracted the background using a polynomial function up to the fifth order and extracted the oscillation components [Figs. S1(b) and S1(d)]. Then, we carried out the well-established pretreatment procedure for Fourier transform of quantum oscillations [S5]. First, we interpolated the background-subtracted data to prepare an equally spaced data set as a function of $1/(\mu_0H)$. Then, we multiplied the Hanning window function to obtain the periodicity of the experimental data. Finally, we performed fast Fourier transform on the treated data set.

At 2 K and 3 T < $\mu_0H$ < 9 T, SdH oscillations with low frequencies, whose frequency (25-32 T) corresponds to that of Weyl fermions in SrRuO$_3$, were observed [Figs. S1(b) and S1(c)]. In addition to the SdH oscillations having low frequencies (25-32 T), we observed SdH oscillations from electrons in the trivial Ru 4$d$ bands having high frequencies (360 and 3400-3800 T) at 100 mK and 12 T < $\mu_0H$ < 14 T [Figs. S1(d) and S1(e)]. Since the carriers in the trivial orbits in SrRuO$_3$ have larger cyclotron masses than those of the Weyl fermions [S2-S4], measurements of the former oscillations should be carried out in relatively low-$T$ and high-$\mu_0H$ regions. Details of these oscillations with low and high frequencies were described in our previous study [S2].

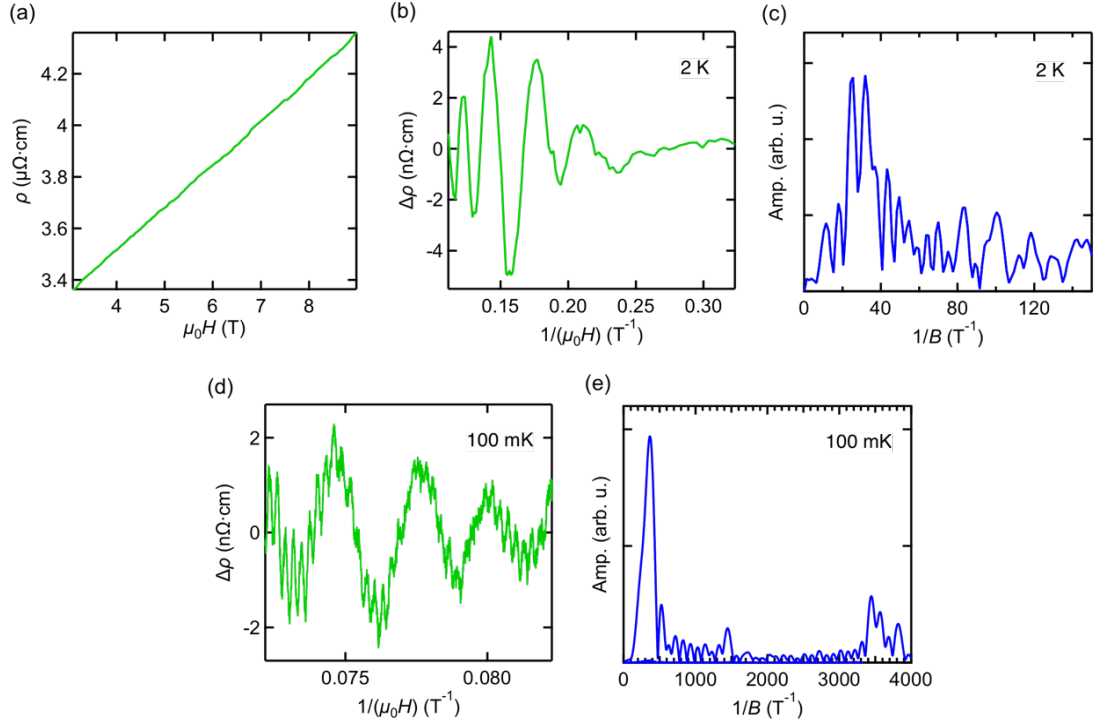

FIG. S1. (a) MR $(\rho(\mu_0 H) - \rho(0\text{ T}))/\rho(0\text{ T})$ at 2 K with $\mu_0 H$ (3 T $< \mu_0 H <$ 9 T) applied in the out-of-plane [001] direction of the $SrTiO_3$ substrate for the $SrRuO_3$ film. (b) SdH oscillations at 2 K with $\mu_0 H$ (3 T $< \mu_0 H <$ 9 T). (c) Fourier transform spectra of the SdH oscillations in (b). (d) SdH oscillations at 100 mK with $\mu_0 H$ (12 T $< \mu_0 H <$ 14 T). (e) Fourier transform spectra of the SdH oscillations in (d).